| | |
|---|---|
| Title | **How do the barrier thickness and dielectric material influence the filamentary mode and CO$_2$ conversion in a flowing DBD?** |
| Authors | A Ozkan[1,2], T Dufour[1], A Bogaerts[2] and F Reniers[1] |
| Affiliations | 1 Université Libre de Bruxelles, Chimie Analytique et Chimie des Interfaces (CHANI), Campus de la Plaine, Bâtiment A, CP255, Boulevard du Triomphe, 1050 Bruxelles, Belgium<br>2 Universiteit Antwerpen, Research Group PLASMANT, Universiteitsplein 1, 2610 Antwerpen-Wilrijk, Belgium |
| Ref. | Plasma Sources Sci. Technol. 25 (2016) 045016 (11pp) |
| DOI | doi:10.1088/0963-0252/25/4/045016 |
| Abstract | Dielectric barrier discharges (DBDs) are commonly used to generate cold plasmas at atmospheric pressure. Whatever their configuration (tubular or planar), the presence of a dielectric barrier is mandatory to prevent too much charge build up in the plasma and the formation of a thermal arc. In this article, the role of the barrier thickness (2.0, 2.4 and 2.8 mm) and of the kind of dielectric material (alumina, mullite, pyrex, quartz) is investigated on the filamentary behavior in the plasma and on the CO$_2$ conversion in a tubular flowing DBD, by means of mass spectrometry measurements correlated with electrical characterization and IR imaging. Increasing the barrier thickness decreases the capacitance, while preserving the electrical charge. As a result, the voltage over the dielectric increases and a larger number of microdischarges is generated, which enhances the CO$_2$ conversion. Furthermore, changing the dielectric material of the barrier, while keeping the same geometry and dimensions, also affects the CO$_2$ conversion. The highest CO$_2$ conversion and energy efficiency are obtained for quartz and alumina, thus not following the trend of the relative permittivity. From the electrical characterization, we clearly demonstrate that the most important parameters are the somewhat higher effective plasma voltage (yielding a somewhat higher electric field and electron energy in the plasma) for quartz, as well as the higher plasma current (and thus larger electron density) and the larger number of microdischarge filaments (mainly for alumina, but also for quartz). The latter could be correlated to the higher surface roughness for alumina and to the higher voltage over the dielectric for quartz. |

# Introduction

Since the nineteenth century, the greenhouse gas concentrations have constantly increased, mainly due to anthropogenic activities using fossil fuels: coal, petroleum products and natural gas [1–4]. In recent decades, the significant amounts of carbon dioxide released into the atmosphere are considered to be responsible for the global warming [5–7]. Spurred by worries on climate changes, the European Commission has implemented increasingly stricter limits on CO$_2$ emissions and has run the H2020 work program with a societal challenge, clearly identified as climate action and environment. More generally, an increasing number of countries become aware of these issues: as an illustration, the COP-21 (Conference of the parties) in 2015 has gathered almost 200 countries around a project agreement on the climate.

Besides the policies for enhancing the energy efficiency and the development of renewable energy, CO$_2$ capturing turns out as a promising alternative through two solutions: (i) geological storage in deep underground [8–10] and (ii) CO$_2$ valorization. The philosophy of the latter is to consider CO$_2$ as a feedstock and not as a waste [11]: CO$_2$ can be converted into value-added products such as carbon monoxide, which is more reactive than CO$_2$. Carbon monoxide is also utilized in several industrial processes, such as Fischer–Tropsch, in a syngas mixture (CO/H$_2$) to produce hydrocarbons [12]. As the dissociation of inert CO$_2$ into CO requires energy, cold atmospheric plasma processes appear as a convenient and innovative method, since most of the energy is supplied by electrons, avoiding excessive energy losses in gas heating. The literature mentions several non-equilibrium plasma sources already in use for CO$_2$ splitting, e.g. gliding arcs [13–17], microwave plasmas [18–20] and dielectric barrier discharges (DBDs) [21–26]. A DBD reactor shows some advantages, e.g. it is easy-to-handle, it operates at atmospheric pressure, it is easy for up-scaling and for combination with catalysis. However, the energy efficiency is still too limited. The average electron energy is typically between 1 and 5 eV [27–31], depending on power, frequency, the nature of the flowing gas and dielectric barrier characteristics. Indeed, every gas behaves differently in a discharge since the reactions with electrons and their energy dependence are different. Ar, He and $N_2$ are already known to change the $CO_2$ discharge behavior by changing for example the density and energy of electrons [32–35].

Since the C = O bond dissociation energy of the $CO_2$ molecule is 5.52 eV, electrons must have energies larger than this value to directly dissociate $CO_2$. Typically this corresponds to the tail of the electron energy distribution function (EEDF) and only a small fraction of electrons has such high energies. Most electrons have energies around 2–3 eV [36], which is also somewhat too high for exciting the vibrational states of $CO_2$. Indeed, Aerts et al [36] and Kozák et al [37] have shown that these vibrationally excited states have a minor influence on the $CO_2$ splitting in a DBD. As vibration-induced dissociation is considered the most energy-efficient process for $CO_2$ splitting, this explains the current limited energy efficiency of a DBD for $CO_2$ splitting.

A packed bed DBD typically yields a higher conversion and energy efficiency [24, 38–41]. For example, packing a DBD reactor with dielectric zirconia ($ZrO_2$) beads enhances the $CO_2$ conversion and energy efficiency by a factor of 1.9 and 2.2, respectively, reaching a conversion up to 42% for a flow rate of 20 mL · min$^{-1}$, and a maximum energy efficiency of 10% at a flow rate of 100 mL · min$^{-1}$ [35]. These improvements are attributed to polarization of the dielectric beads, enhancing the local electric field. Furthermore, the material of the dielectric barrier is also of great importance [24, 42–44]. Very few experimental papers explain how the barrier thickness of a DBD can influence the filamentary behavior of the DBD, especially on the topic of gas treatment [45–50]. To the authors' knowledge, only Forte et al have studied this effect. They consider that reducing the thickness makes the discharge more unstable and large energetic filaments can appear and damage the barrier due to strong local heating.

In this work, a flowing DBD source at atmospheric pressure is used to split CO2 into CO, O and O2. We investigate the effect of the dielectric barrier on the $CO_2$ conversion and energy efficiency, by varying its thickness as well as its material. We characterize the filamentary mode of the discharge to elucidate the role played by the microdischarges on the $CO_2$ conversion. For that purpose, a detailed electrical characterization is performed to obtain information on the number and lifetime of the microdischarges, the plasma current and electrical charge.

## Experimental set-up

### Plasma reactor and set of dielectric barriers

The tubular DBD reactor is shown in figure 1. The central cylindrical copper electrode is 22 mm in diameter and 120 mm in length. It is powered by a high AC voltage, whereas the outer electrode is grounded. The latter is a stainless steel mesh, 100 mm long and rolled around the tubular dielectric barrier. This barrier always has an inner diameter of 26 mm to fix the electrode-barrier gap at 2 mm in all experiments, and therefore keeping the same discharge volume of 15.0 cm$^3$. Since the $CO_2$ flow rate is set at 200 mL$_n$ · min$^{-1}$ the residence time is estimated to 4.5 s. The applied power is provided by an AFS generator G10S-V coupled with a transformer.

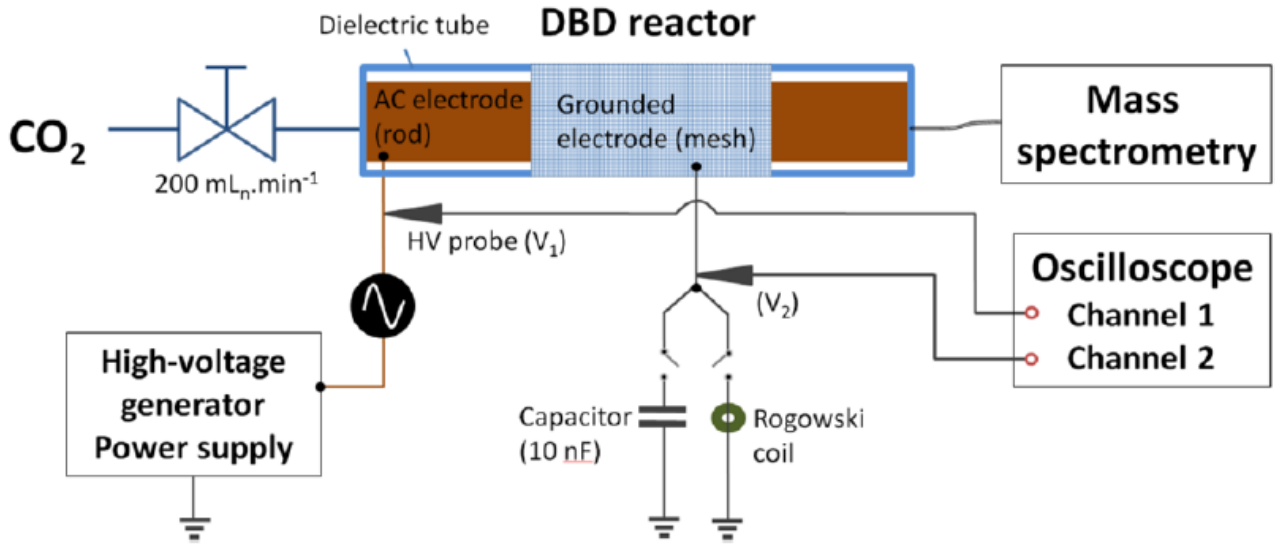

*Figure 1. Schematic diagram of the experimental set-up.*

To study the effect of the dielectric thickness, we used pyrex (borosilicate glass 3.3) with 3 different thicknesses: 2.0, 2.4 and 2.8 mm (±0.02 mm). As mentioned above, the inner diameter is fixed while the outer diameter is changed. Furthermore, four different dielectric barriers with the same thickness (2 mm) but different relative permittivities ($\varepsilon_r$) are compared: alumina ($\varepsilon_r$ = 9.6), mullite ($\varepsilon_r$ = 6.0), pyrex ($\varepsilon_r$ = 4.6) and quartz ($\varepsilon_r$ = 3.8). These $\varepsilon_r$ values remain constant for our conditions of frequency (between 1 kHz and 1 MHz) and temperature (from 300 K to 473 K for quartz and 573 K for alumina) [51, 52]. Additional properties of the barriers are indicated in table 1, i.e. the chemical composition, the surface roughness measured by profilometry, the thermal conductivity and the capacitance. The latter parameter is calculated considering the following equation, where L, $r_{in}$ and $r_{out}$ are the dielectric length, the inner and outer radius, respectively:

$$C = \frac{2\pi\varepsilon_0\varepsilon_r \cdot L}{\ln(r_{out}/r_{in})}. \quad (1)$$

| Material | Chemical composition | Relative permittivity ($\varepsilon_r$) | $R_{RMS}$ (nm) | Thermal conductivity (W·m$^{-1}$·K$^{-1}$) at 20 °C | Capacitance (pF) |
|---|---|---|---|---|---|
| Alumina (C799) | Al$_2$O$_3$ 99.70% Na$_2$O 0.15% SiO$_2$ 0.10% | 9.6 | 6800 | 29 | 373.2 |
| Mullite (C610) | Al$_2$O$_3$ 62.60% SiO$_2$ 35.15% Fe$_2$O$_3$ 0.82% TiO$_2$ 0.39% CaO 0.18% | 6.0 | 3100 | 2 | 233.2 |
| Pyrex (3.3 Duran) | SiO$_2$ 81% B$_2$O$_3$ 13% Na$_2$O + K$_2$O 4% Al$_2$O$_3$ 2% | 4.6 | 780 | 1.1 | 178.8 |
| Quartz | SiO$_2$ (high purity) | 3.8 | 89 | 1.4 | 147.7 |

*Table 1. Physical properties of the four different dielectric barriers tested, 2 mm in thickness*

| Parameter | Formula | # |
|---|---|---|
| Conversion | $CO_2\,(\%) = \dfrac{I_{CO_2}^{plasma\,OFF} - I_{CO_2}^{plasma\,ON}}{I_{CO_2}^{plasma\,OFF}} \times 100\,\%$ | Equation (2) |
| Energy efficiency | $\eta_{CO_2}\,(\%) = \chi_{CO_2}(\%) \cdot \dfrac{\Delta H^0_{298\,K\,(eV\cdot molecule^{-1})}}{SEI_{(eV\cdot molecule^{-1})}}$ | Equation (3) |
| Energy density | $E_{d\,(J\cdot cm^{-3})} = \dfrac{\text{Absorbed power}_{(J\cdot s^{-1})}}{\text{Gas flow rate}_{(cm^3\cdot s^{-1})}}$ | Equation (4) |
| Specific energy input | $SEI_{(eV\cdot mol^{-1})} = \dfrac{E_{d(J\cdot cm^{-3})} \times 6.24 \times 10^{18}{}_{(eV\cdot J^{-1})} \times 24\,500_{(cm^3\cdot mol^{-1})}}{6.022 \times 10^{23}{}_{(molecule\cdot mol^{-1})}}$ | Equation (5) |

*Table 2. Formulas for conversion, energy efficiency, energy density and specific energy input.*

## Mass spectrometry (MS)

After passing through the reactor, the gas is analyzed by a mass spectrometer operating at atmospheric pressure with a quadrupole gas analyzer (Hiden Analytical QGA, Warrington, UK). The electron energy in the ionization chamber is set at 70 eV and the detector is a secondary electron multiplier (SEM). MASsoft7 software is used to monitor simultaneously the partial pressure variations with specific m/z ratios as a function of time. The $CO_2$ conversion ($CO_2$) is calculated according to equation (2) in table 2, where I corresponds to the $CO_2$ intensity signals in the mass spectrometer. The energy efficiency of the $CO_2$ conversion is calculated from $CO_2$, the enthalpy of the splitting reaction ($CO_2 \rightarrow CO + ½O_2$), namely $\Delta H°_{298K}$ = 282.99 kJ · mol$^{-1}$ = 2.94 eV · molecule$^{-1}$ and the specific energy input (SEI) (see table 2). Note that the enthalpy of the reaction almost does not change in a temperature range from 298 to 473 K, which is a typical gas temperature inside the DBD plasma [26, 53]. The energy density and the specific energy input are also presented in table 2.

## Electrical measurements

The electrical measurements are performed with a Tektronix DPO 3032 oscilloscope and a Tektronix P6015A probe. According to figure 1, the voltage supplying the plasma source (VDBD) is expressed as the difference of potentials V1 and V2, but also as the sum of two voltages: the dielectric voltage ($V_{diel}$) and the effective plasma voltage ($V_{pl,eff}$). The plasma voltage is considered as effective since the filamentary mode is responsible for an inhomogeneous electric field in the whole electrode-barrier gap, which is different from the case of a diffuse and homogeneous glow discharge [29, 54–56]. Therefore, $V_{pl,eff}$ should be considered as an average value and represents typically 70% of VDBD [26]. As indicated in figure 1, the potential $V_2$ is measured either through a capacitor to evaluate the power absorbed by the plasma ($P_{abs}$) via the Lissajous method [57, 58], or through a current probe (Pearson 2877 Rogowski coil), both placed in series with the DBD.

An atmospheric $CO_2$ plasma generated in a flowing DBD typically operates in the filamentary mode [21, 36, 59]. Therefore, the discharge current presents two components: the dielectric current (a sinusoidal-like signal) and the plasma current (peaks superposed to the previous signal and representing the microdischarges). Based on a numerical method validated in our previous work [26], the microdischarges are investigated through their individual features, such as average lifetime ($L_{md}$) and electrical charge, but also through their collective features, such as the plasma charge accumulation and their total number ($N_{md}$) for a given analysis time. These data are collected for 20 periods—which corresponds to an analysis time of 700 ms—and then averaged over a single period in order to have statistically meaningful results. All this information is of great importance for modeling $CO_2$ conversion in a DBD in filamentary mode [36, 60].

### Infrared imaging

2D temperature profiles of the grounded outer electrode and of the reactor wall are measured with an infrared camera (FLIR E40) with a resolution of 160 × 120 pixels and a thermal sensitivity lower than 0.07 °C at 30 °C. FLIR ResearchIR software is used to control, record and analyze the temperature profiles in a range from −20 °C to +650 °C. The emissivity coefficients are introduced in the software. The temperature is calibrated at room temperature.

### Profilometry

Profilometry measurements on the dielectric surfaces are performed using a Brücker dektak XT stylus profilomètre (Brüker, Karlsruhe, Germany). The scanning stylus is 2 µm in radius and is applied with a force of 1 mg. The roughness, i.e. $R_{RMS}$ parameter, is estimated using the Vision 64 software by summing 150 scans over a 0.9 mm$^2$ area.

## Results and discussion

### Effect of the barrier thickness

**$CO_2$ conversion and energy efficiency**

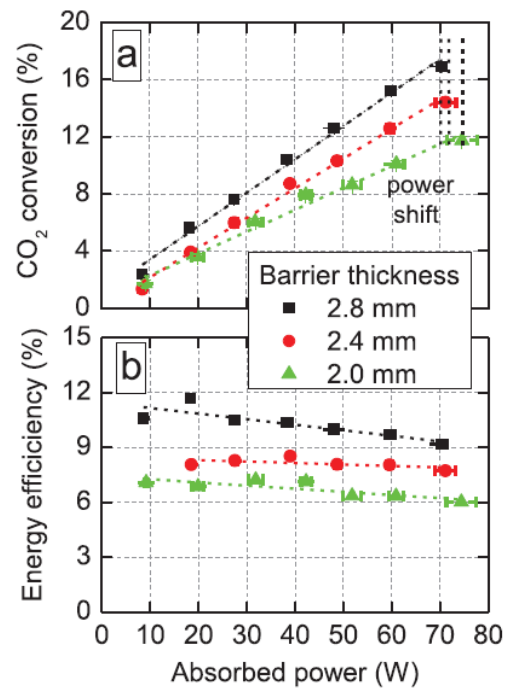

Figure 2 presents the $CO_2$ conversion versus the absorbed power, for different barrier thicknesses. This figure clearly shows that increasing the absorbed power improves the $CO_2$ conversion. More electrons are indeed produced and likely to participate to the splitting process. Usually, this rise in conversion is correlated with a drop in the energy efficiency [59]. However, in our case this drop is very minor because the conversion rises almost proportionally with the absorbed power (see equations (3)–(5)). The highest conversion (17%) and energy efficiency (9%) are obtained for an absorbed power of 75 W (i.e. specific energy input of 5.75 eV · molecule$^{-1}$) and the largest dielectric thickness (2.8 mm). These values are in line with typical values found in literature. Indeed, the maximum energy efficiency of a DBD in pure $CO_2$ is usually comprised between 3 and 9%, with maximum conversions reported between 13 and 35% [24, 59, 61].

*Figure 2. (a) $CO_2$ conversion and (b) energy efficiency as a function of the absorbed (plasma) power for three different dielectric thicknesses (2.0, 2.4 and 2.8 mm); f = 28.6 kHz; $\Phi(CO_2)$ = 200 mL$_n$ · min$^{-1}$.*

Furthermore, figure 2 indicates that for a fixed absorbed power, the conversion and energy efficiency always increase with rising thickness of the barrier. For instance, at 50 W, an enhancement of 50% is obtained when increasing the thickness from 2.0 mm to 2.8 mm. Also, for measurements performed at fixed applied power, the data points show a small horizontal shift because a larger barrier thickness induces a higher reflected power (and hence a slightly lower absorbed power). To understand how a thicker barrier improves the $CO_2$ conversion and energy efficiency, a detailed electrical characterization is presented in the next section.

**Electrical characterization**

As shown in figure 3(a), the voltage applied to the DBD reactor ($V_{DBD}$) increases with the absorbed power. This was also reported in [26]. At fixed power, figure 3(b) illustrates that VDBD can significantly rise with the dielectric thickness, e.g. at 60 W it linearly increases from 5050 V to 5600 V (RMS values) for a barrier thickness ranging from 2.0 to 2.8 mm. As mentioned before, $V_{DBD}$ consists of two components—averaged plasma voltage ($V_{pl,eff}$) and dielectric voltage ($V_{diel}$). It is clear from figure 3(b) that the rise in $V_{DBD}$ is attributed to $V_{diel}$, while the plasma voltage remains constant and close to 3800 V. This means that the electric field remains constant whatever the barrier thickness and therefore, this cannot explain the rise in $CO_2$ conversion.

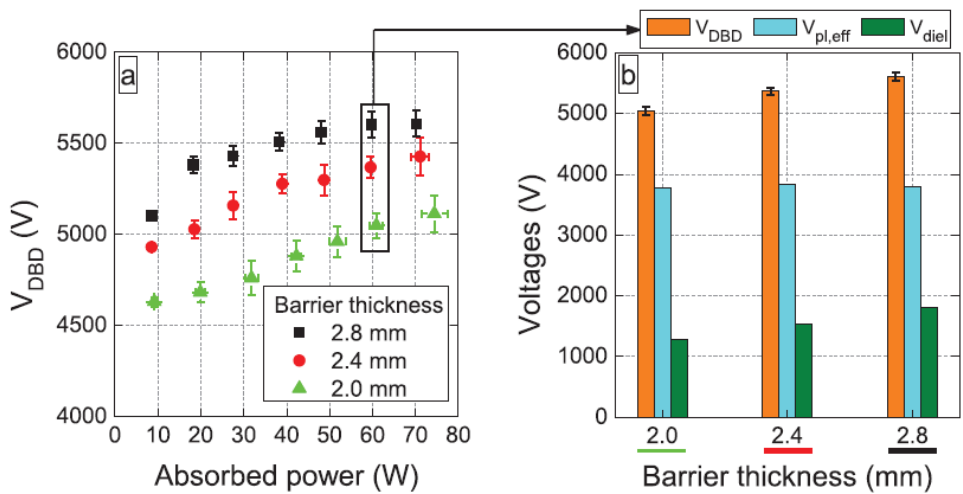

*Figure 3. (a) Applied voltage (VDBD) as a function of absorbed power for three different dielectric thicknesses, and (b) voltage components as a function of the dielectric thickness at a given absorbed power $P_{abs}$ = 60 W; f = 28.6 kHz; $\Phi(CO_2)$ = 200 $mL_n \cdot min^{-1}$.*

However, the latter can be explained by performing a detailed analysis of the currents to extract information about the microdischarges, i.e. their individual features such as their average lifetime and electrical charge, but also their collective features such as the plasma charge accumulation and their total number for a given analysis time (e.g. period or residence time). Table 3 summarizes the average number ($N_{md}$) and lifetime ($L_{md}$) of the microdischarges for one period, as a function of the barrier thickness. Increasing the barrier from 2.0 to 2.8 mm leads to a significant increase of $N_{md}$ (from 465 to 506) and a slight decrease in $L_{md}$ (from 13.3 ns to 12.3 ns). However, the electrical charge accumulated on the barrier remains unchanged and close to 1 µC (see $Q_{diel}$ in table 3). The same applies to the charge accumulated in the plasma ($Q_{plasma}$) and thus to the total charge ($Q_{total}$). For the dielectric, the following equation can thus be written:

$$Q_{diel} = C_{diel} \cdot V_{diel} = \text{Constant, whatever barrier thickness.} \quad (6)$$

According to equation (1), increasing the barrier thickness corresponds to an increase of rout and thus to a drop in the barrier capacitance. Since our measurements reveal that $Q_{diel}$ (=$C_{diel} \cdot V_{diel}$) is constant, an increase of $V_{diel}$ is needed, as confirmed by our results presented in figure 3(b). In other words, the drop in barrier capacitance yields a higher voltage over the dielectric ($V_{diel}$), and the latter causes clearly a larger number of microdischarges per period ($N_{md}$), as reported in table 3.

| Parameters measured for 1 period | | Dielectric barrier thickness | | |
|---|---|---|---|---|
| | | 2.0 mm | 2.4 mm | 2.8 mm |
| $N_{md}$ (−) | | 465 (±20) | 487 (±23) | 506 (±7) |
| $L_{md}$ (ns) | | 13.3 (±0.4) | 12.9 (±0.4) | 12.3 (±0.2) |
| Charge ($\mu C$) | $Q_{total}$ | 1.27 (±0.01) | 1.27 (±0.02) | 1.30 (±0.02) |
| | $Q_{plasma}$ | 0.29 (±0.01) | 0.30 (±0.04) | 0.29 (±0.01) |
| | $Q_{diel}$ | 0.98 | 0.97 | 1.01 |

*Table 3. Number of microdischarges during one period, mean lifetime of the microdischarges and charge accumulation as a function of the dielectric thickness; $P_{abs}$ = 60 W; f = 28.6 kHz; $\Phi(CO_2)$ = 200 $mL_n \cdot min^{-1}$.*

Figure 4 indeed illustrates a higher density of microdischarges upon rising the barrier thickness, always maintaining a uniform spatial distribution in the entire discharge region. This increase is about 9% for a barrier thickness increasing from 2.0 to 2.8 mm. As the reactor volume is the same, independent of the barrier thickness, the probability for a single $CO_2$ molecule to pass through the discharge and interact with at least one microdischarge therefore increases for the thicker barriers. As a result, a higher $CO_2$ conversion (and thus energy efficiency) is obtained. It should be mentioned that the average lifetime of the microdischarges slightly drops upon increasing barrier thickness, but this seems of lower importance for determining the $CO_2$ conversion.

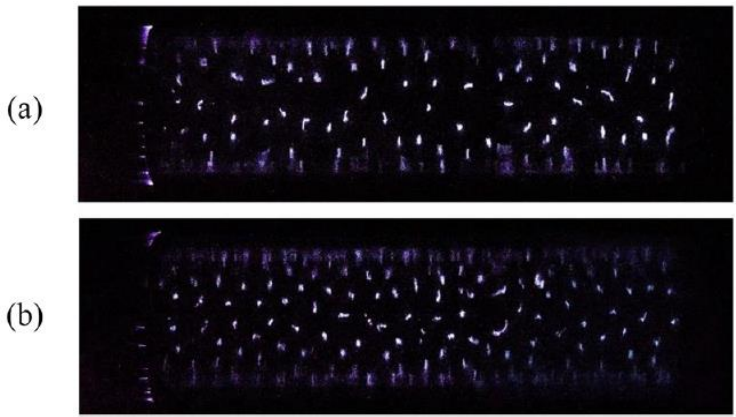

*Figure 4. Pictures of the microdischarges observed through the pyrex and the outer mesh electrode, at the same power, frequency and flow rate (for a camera aperture of 1/100 s). The barrier thickness is (a) 2.0 mm, (b) 2.8 mm.*

## Effect of the dielectric material

**$CO_2$ conversion and energy efficiency**

Figure 5 shows the $CO_2$ conversion and energy efficiency for different dielectric materials, but keeping the same operating conditions and configuration (f = 27.1 kHz; $\Phi(CO_2)$ = 200 $mL_n \cdot min^{-1}$). These materials are indicated on the X-axis, in decreasing order of their relative permittivities (see table 1 above). Each barrier is 2 mm thick.

Surprisingly, the highest $CO_2$ conversions (e.g. 24.6% at 74 W) are obtained for dielectric barriers with the highest and lowest relative permittivities, i.e. alumina and quartz, respectively. Similar results, i.e. the same $CO_2$ conversion for alumina and quartz, were also obtained in [59]. The two other dielectric barriers with intermediate relative permittivities, i.e. mullite and pyrex, yield somewhat lower $CO_2$ conversion. The energy efficiency follows the same trend, and here the results are even somewhat higher for quartz (i.e. above 15% at 39 W).

At first sight, the results presented in figure 5 seem counterintuitive, if we only consider a change in the $\varepsilon_r$ parameter (i.e. capacitance). However, changing the nature of the dielectric material does not solely mean a change of its capacitance. Other relevant parameters of the barrier can also influence the $CO_2$ conversion in a DBD, such as its surface roughness and thermal conductivity, as will be explained below. However, first we present a detailed electrical characterization, as this will clarify the observed trends in $CO_2$ conversion and energy efficiency.

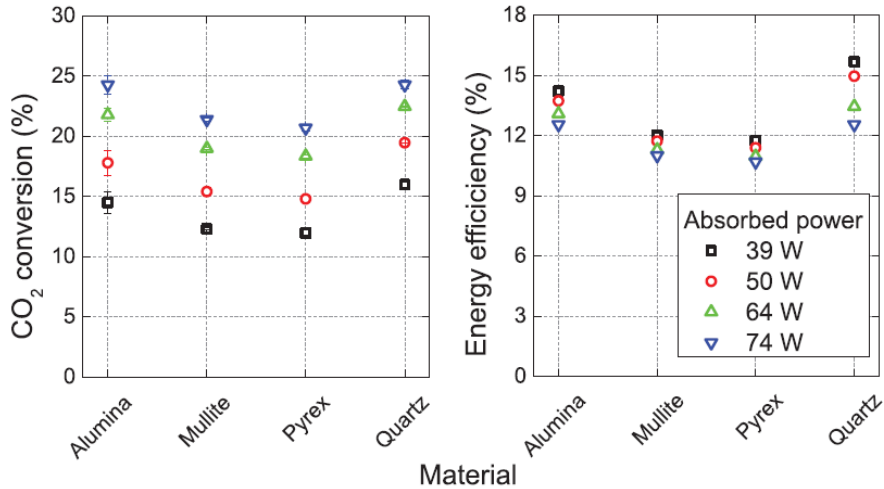

*Figure 5. (a) $CO_2$ conversion and (b) energy efficiency for different values of the absorbed power and for four different dielectric materials; f = 27.1 kHz; $\Phi(CO_2)$ = 200 mL$_n$ · min$^{-1}$.*

**Electrical characterization**

$V_{DBD}$ is plotted as a function of the absorbed power in figure 6(a), for the four different dielectric materials. $V_{DBD}$ clearly rises with increasing power, but also with decreasing relative permittivity of the materials. The latter is also shown in figure 6(b), for a fixed power of 75 W. Moreover, both the average plasma voltage ($V_{pl,eff}$) and the voltage over the dielectric ($V_{diel}$) slightly rise upon decreasing relative permittivity, although pyrex is behaving somewhat differently. Hence, this behavior might partially explain the higher $CO_2$ conversion and energy efficiency in the case of quartz, as a higher average plasma voltage yields a higher electric field in the gap, which results in more electron heating and hence, in a higher $CO_2$ conversion by electron impact dissociation. However, this behavior does not explain the higher conversion and energy efficiency for alumina compared to mullite and pyrex. Therefore, there must be other effects coming into play as well. In order to explain this, the relation between the trend in $CO_2$ conversion and the specific properties of the microdischarges will be discussed in the next paragraph.

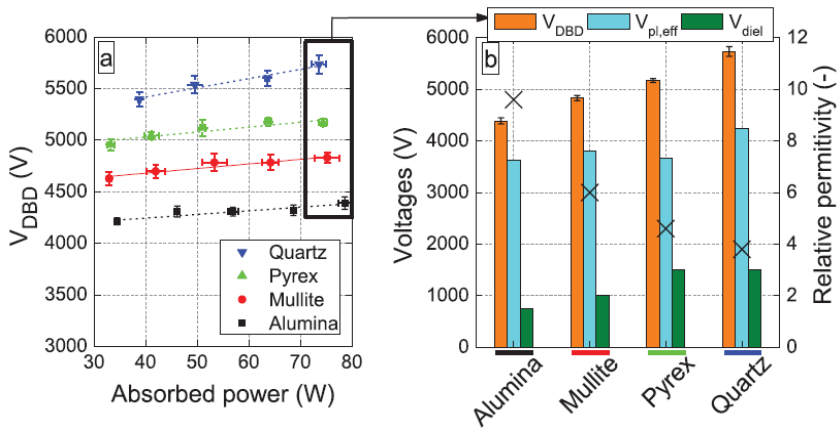

*Figure 6. (a) RMS applied voltage ($V_{DBD}$) as a function of absorbed power for four different dielectric materials at f = 27 kHz, and (b) RMS voltage components as a function of the dielectric material at a given absorbed power $P_{abs}$ = 75 W; $\Phi(CO_2)$ = 200 mL$_n$ · min$^{-1}$. The relative permittivities of the materials are indicated with crosses, referring to the right y-axis.*

The oscillograms of the total current in a DBD with alumina, mullite, pyrex and quartz are plotted in figure 7. By comparing these current profiles, one can observe that the discharge filamentation is different for the different dielectric materials, and the peak distribution in the positive and negative half cycle during one period is also different for the different materials. Usually—and as evidenced above for the effect of the barrier thickness—decreasing the capacitance leads to an increase in the voltage over the dielectric (see also figure 6), so that the plasma charge remains unchanged. However, in the present comparison of the different dielectric materials, the plasma charge (or plasma current) is not constant: as shown in figure 8(a), it is much higher in the case of alumina than for the other dielectric materials. Likewise, the number of microdischarge filaments, as well as their average lifetime, is higher for alumina as well (see figures 8(b) and (c)). The alumina barrier gives rise to 430–570 microdischarges over one period (depending on the power), while this value varies between 350 and 490 for the other materials. Moreover, the mean lifetime of the microdischarges is also longer for alumina—between 11.5 and 15.3 ns, depending on power— while it is between 10.5 and 13.5 ns for the other materials.

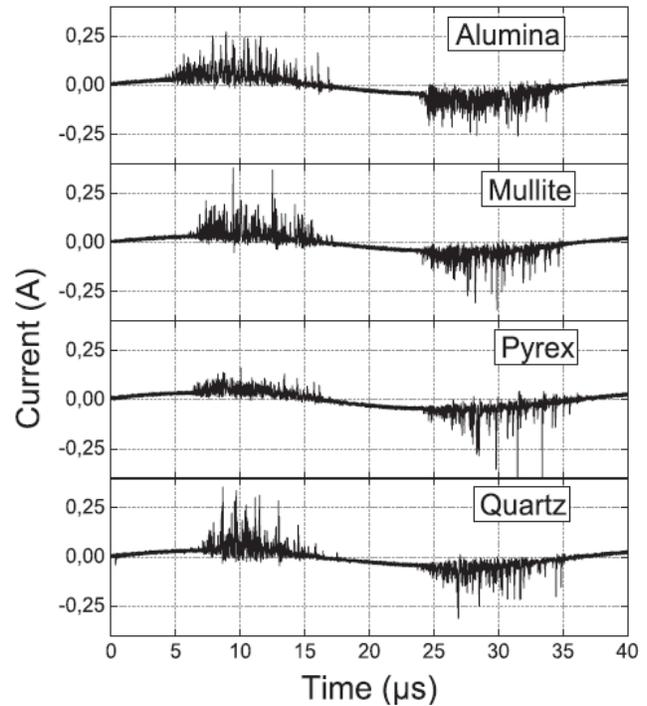

*Figure 7. Current profiles as a function of time for four different dielectric materials; $P_{abs}$ = 75 W; $\Phi(CO_2)$ = 200 $mL_n \cdot min^{-1}$.*

Alumina thus gives rise to a filamentary discharge with more microfilaments, which are also broader compared to the other discharges. As the discharge zone is the same in all cases, the larger number of filaments (with longer average lifetime) means that a larger discharge volume is available for the $CO_2$ conversion, and the latter can explain why alumina yields a higher $CO_2$ conversion and energy efficiency. Furthermore, alumina gives a higher plasma charge, and this implies a higher electron density, which might also explain the higher $CO_2$ conversion and energy efficiency, for the same power as for the other materials.

In summary, the trends of the plasma charge (figure 8(a)) (which reflects the plasma current and hence, the number of electrons present in the discharge region), the number of microdischarge filaments (figure 8(b)) and average lifetime (figure 8(c)), in combination with the somewhat higher effective plasma voltage (and thus the somewhat higher electric field and electron energy in the plasma) for quartz (figure 6(b)), can clearly explain the trend of the $CO_2$ conversion and energy efficiency, as observed in figures 5(a) and (b).

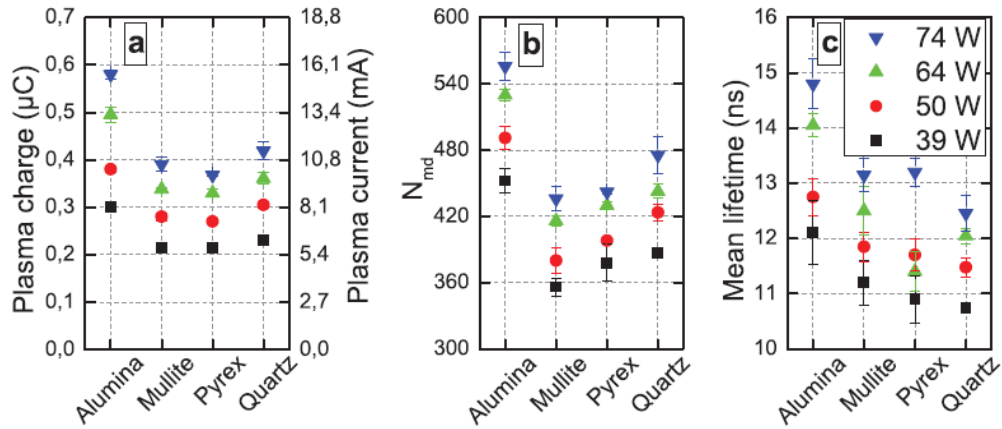

*Figure 8. (a) Plasma charge accumulation during one period, (b) number of microdischarges during one period and (c) mean lifetime of the microdischarges, for the four different dielectric materials and for different absorbed powers and $\Phi(CO_2) = 200$ mL$_n \cdot$ min$^{-1}$.*

**Wall reactor temperatures and roughness analysis of the dielectric material**

To explain why alumina yields a larger number of microdischarge filaments than the other materials, we have investigated the surface roughness of the dielectric materials. A higher surface roughness can indeed imply easier generation of filaments. The roughness was estimated through the RRMS parameter and is reported in table 1 above for each material. It is clear that alumina has by far the largest roughness (in average 6800 nm), and this could explain why alumina induces more microdischarges, as observed above (cf figure 8(b)). On the other hand, the surface roughness does not explain why quartz also induces a relatively large number of microdischarges, as it has a clearly lower surface roughness than mullite and pyrex. In this case, it can be attributed to the higher voltage over the dielectric, as illustrated in figure 6, exactly as for the effect of the barrier thickness (see previous section). In summary, the larger number of filaments can be attributed to the higher voltage over the dielectric (like in the case of quartz, and for the effect of the dielectric thickness—higher at 2.8 mm of thickness) or to the higher surface roughness (like in the case of alumina).

Finally, the last material property that we have investigated is the thermal conductivity. As indicated in table 1, the thermal conductivity of alumina is (more than) ten times higher than for the three other materials. Hence, when the plasma is ignited, the heating dissipation through the alumina barrier is much faster, yielding a lower wall temperature as compared to the other materials. In figure 9, the average temperature of the mesh outer electrode is 136 °C in case of alumina, against 149 °C, 157 °C and 169 °C in case of mullite, quartz and pyrex, respectively. Moreover, the wall temperature outside of the discharge zone is quite elevated for alumina (almost 80°C) while it is only 45 °C for the three other materials. Thus in the latter cases, the heating appears clearly confined in the discharge region. The different surface temperature might also yield a different gas temperature (inside the discharge region, as well as before or after), and this may also affect the $CO_2$ conversion, due to changes in the chemical reaction rates at different temperature. The rate constants of the heavy particle reactions are indeed often a function of the gas temperature. The temperature-dependence of the rate constants of the neutral reactions and of some ion reactions in a $CO_2$ DBD plasma is presented in [36, 37, 59, 62]. Kozak et al evaluated the effect of the gas temperature on the $CO_2$ conversion and energy efficiency by means of plasma chemistry modeling [62]. We expect that the temperature affects the $CO_2$ conversion much more in a microwave plasma [62] than in a DBD plasma [59], because of the important role of the vibrational kinetics in a MW plasma, and because the latter are very much temperature-dependent [37, 62]. Moreover, besides the effect of the gas temperature on the reaction rate constants, it also affects the particle densities through the ideal gas law, i.e. at constant pressure, a higher temperature yields a lower gas density. As a result, the reduced electric field (E/N) will rise, and this will affect the electron energy distribution function (EEDF), which will in turn affect the electron impact reaction rates, and thus the $CO_2$ conversion. However, the exact reason why the changes in gas temperature might affect the $CO_2$ conversion is beyond the scope of this paper.

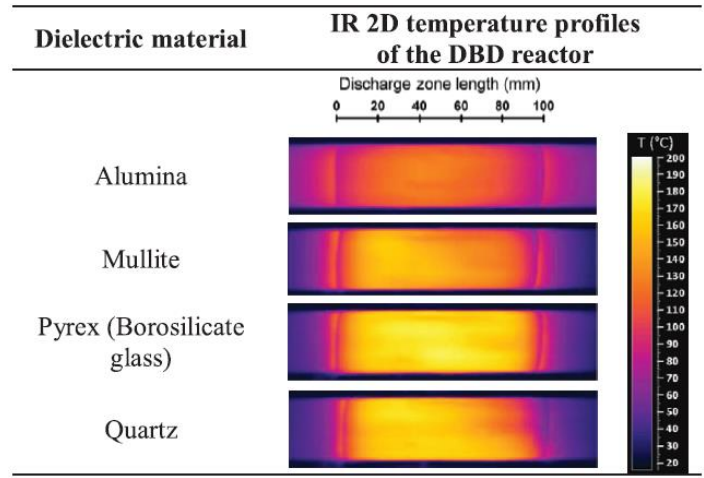

*Figure 9. 2D temperature profiles of the DBD reactors for the four different dielectric materials, for $P_{abs}$ = 75 W, process time = 4 min. The black graded line indicates the length of the outer electrode, i.e. the discharge region.*

# Conclusion

We have investigated how the thickness and the material of the dielectric barrier affect the $CO_2$ conversion in a DBD reactor operating at atmospheric pressure. Table 4 summarizes the results of all the experiments with arrows (for the effect of barrier thickness) and plus signs (for the different materials), to correlate the $CO_2$ conversion and energy efficiency with the other plasma characteristics (like the voltage, the microdischarge properties and the charges in the plasma) and with the material characteristics (like thermal conductivity and surface roughness). From this correlation, the trends in $CO_2$ conversion (and energy efficiency) can clearly be explained.

|  |  | Increasing the dielectric barrier thickness (from 2 to 2.8 mm) | Dielectric materials | | | |
|---|---|---|---|---|---|---|
|  |  |  | Quartz | Pyrex | Mullite | Alumina |
| $CO_2$ conversion ($\chi_{CO_2}$) | | ↗ | ++++ | ++ | + | ++++ |
| Energy efficiency ($\eta_{CO_2}$) | | ↗ | ++++ | ++ | + | ++++ |
| Voltage | Applied ($V_{DBD}$) | ↗ | ++++ | +++ | ++ | + |
|  | Plasma ($V_{pl,eff}$) | = | ++++ | + | ++ | + |
|  | Dielectric ($V_{diel}$) | ↗ | ++++ | ++++ | ++ | + |
| Microdischarges | Number ($N_{md}$) | ↗ | ++ | + | + | ++++ |
|  | Lifetime ($L_{md}$) | ↘ | + | ++ | ++ | ++++ |
|  | Current ($i_{pl}$) | = | ++ | + | + | ++++ |
| Charge | Plasma ($Q_{pl}$) | = | ++ | + | + | ++++ |
|  | DBD ($Q_{DBD}$) | = | / | / | / | / |
| Energy loss to Joule effect (thermal conductivity) | | ↘ | ++++ | ++++ | +++ | + |
| Roughness of dielectric material | | = | + | + | +++ | ++++ |

*Table 4. Summary of $CO_2$ conversion and energy efficiency as a function of the barrier thickness and kind of dielectric material, and correlation with the plasma and material characteristics.*

By increasing the thickness of the barrier from 2.0 to 2.8 mm, the $CO_2$ conversion (and thus also the energy efficiency) clearly increases by about 50%. The best results, in terms of both conversion and energy efficiency, are obtained at an absorbed power of 70 W (corresponding to a SEI of 5.75 eV · molecule$^{-1}$) and the largest dielectric thickness (2.8 mm), yielding a conversion of 17% and a corresponding energy efficiency of 9%. As indicated in table 4, the charge and the average plasma voltage remain unchanged and therefore, they cannot explain the higher conversion. The reason for the higher $CO_2$ conversion is the larger number of microdischarges in a certain period. Indeed, the $CO_2$ gas flowing through the reactor will have a larger chance to pass through at least one microdischarge, explaining the higher conversion.

Among the four dielectric materials investigated, quartz and alumina lead to the highest $CO_2$ conversion (i.e. 24.6% at 74 W) and energy efficiency (i.e. above 15% at 39 W for quartz, and slightly lower for alumina). This can be explained from the plasma charge, the number of microdischarge filaments and their average lifetime, which are clearly the highest for alumina, and the second highest for quartz. In addition, the high values for quartz, which are a bit counter-intuitive because this material has the lowest relative permittivity, can be explained from the higher effective plasma voltage (and thus the somewhat higher electric field and electron energy in the plasma). In general, the relative permittivity of the materials seems not to be important for determining the $CO_2$ conversion and energy efficiency, at least not in the range investigated in this study (i.e. $\varepsilon_r$ between 3.8 and 9.6). On the other hand, the larger number of microdischarges in a certain period seems mostly responsible for the higher $CO_2$ conversion, just like in the case of the effect of barrier thickness. This larger number of microdischarges could be explained in the case of quartz by the higher voltage over the dielectric (again similar to the effect of the barrier thickness), and in the case of alumina by the higher surface roughness of the material.

# Acknowledgments

The authors acknowledge financial support from the IAPVII/12, P7/34 (Inter-university Attraction Pole) program 'PSI-Physical Chemistry of Plasma-Surface Interactions', financially supported by the Belgian Federal Office for Science Policy (BELSPO). A. Ozkan would like to thank the financial support given by 'Fonds David et Alice Van Buuren'.